\documentclass[manuscript]{aastex}

\usepackage{color}

\shorttitle{Collapse and fragmentation of massive magnetized dense cores using radiation-magneto-hydrodynamics.}
\shortauthors{Commer\c con et al.}

\begin{document}


\title{Collapse  of massive magnetized dense cores using radiation-magneto-hydrodynamics: early fragmentation inhibition.}


\author{Beno\^{i}t Commer\c con\altaffilmark{1}}
\affil{Max Planck Institut f\H{u}r Astronomie, K\H{o}nigsthul 17, 69117 Heidelberg, Germany}
\email{benoit@mpia-hd.mpg.de}

\author{Patrick Hennebelle \altaffilmark{2}}
\affil{Laboratoire de radioastronomie, UMR 8112 du CNRS, \'{E}cole normale sup\'{e}rieure et Observatoire de Paris, 24 rue Lhomond, 75231 Paris Cedex 05, France}

\and

\author{Thomas Henning \altaffilmark{1}}
\affil{Max Planck Institut f\H{u}r Astronomie, K\H{o}nigsthul 17, 69117 Heidelberg, Germany}




\begin{abstract}
We report the results of radiation-magneto-hydrodynamics calculations in the context of high mass star formation, using for the first time a self-consistent model for photon emission (i.e. via thermal emission and in radiative shocks) and with the high resolution necessary to resolve properly magnetic braking effects and radiative shocks on scales $<100$ AU. We investigate the combined effects of magnetic field, turbulence, and radiative transfer on the  early phases of the collapse and the fragmentation of massive dense cores. 
We identify a new mechanism that inhibits initial fragmentation of massive dense cores, where magnetic field and radiative transfer interplay. We show that this interplay becomes stronger as the magnetic field strength increases. Magnetic braking is transporting angular momentum outwards and is lowering the rotational support and is thus increasing the infall velocity. This enhances the radiative feedback owing to the accretion shock on the first core. We speculate that highly magnetized massive dense cores are good candidates for isolated massive star formation, while moderately magnetized massive dense cores are more appropriate to form OB associations or small star clusters. 

\end{abstract}



\section{Introduction}

Massive star formation (M$_\star > 8$ M$_\odot$) is one of the most challenging astrophysical problems. It is established that most massive stars form from massive prestellar cores and occur in high-order multiple systems  \citep[e.g.][]{Zinnecker_Yorke_2007}. Nevertheless, all theoretical numerical models to date show that massive prestellar cores are unlikely to form without first fragmenting into several objects. 
In addition, recent observational work \citep[e.g.][]{Hennemann_et_al_2009,Bontemps_et_al_2010,Longmore_et_al_2011}  suggests that collapsing
massive dense cores are less fragmented than what numerical calculations
produce although the limited observational resolution available precludes a
definitive answer.
For instance, isothermal simulations  \citep[e.g.][]{Bonnell_et_al_2001,Bonnell_Bate_2006}, radiative
calculations  \citep{Krumholz_et_al_2007,Krumholz_et_al_2009}, magnetized ones \citep{Wang_et_al_2010,Hennebelle_et_al_2011,Peters_et_al_2011,Seifried_et_al_2011} and even calculations including radiative ionisation \citep{Peters_et_al_2010},
tend all to form several fragments. Indeed, both radiation and magnetic
fields tend to reduce the number of fragments (e.g. a factor of about 2 for
highly magnetized cores) without suppressing the fragmentation.
To balance this fragmentation issue, \cite{Krumholz_McKee_2008} proposed a column density threshold for massive star formation. However, this model only applies to  massive star formation under certain conditions, and in particular to massive stars, whose formation has been possible thanks to the radiative feedback of low-mass protostars. 


In this Letter, we perform full radiation-magneto-hydrodynamics (RMHD) calculations of massive, turbulent, and magnetized dense cores. This work is an extension towards higher masses of the study by  \cite{Commercon_2010}, who investigated low-mass star formation.

The paper is organized as follows: In section 1 we discuss our numerical method and initial conditions. Our results are presented in section 2. In section 3 we dicsuss a new scenario for massive  star formation and section 4 concludes the paper.

\section{Model}

\subsection{Initial conditions}

Our initial setup is identical to the one used in \cite{Hennebelle_et_al_2011}, except that we do not use a barotropic equation of state. We consider 100 M$_\sun$ spherical dense cores, which are threaded by a magnetic field parallel to the $x$-direction. The initial radius of the sphere is 0.67 pc and the total box-length is  2.76 pc. The density profile is given by $\rho(r)=\rho_c/(1+(r/r_0)^2)$, where $\rho_c=1.4\times 10^{-20}$ g cm$^{-3}$ is the central density, and $r_0\approx0.22$ pc is the extent of the central plateau. We impose a density contrast between the central density and the edge density of 10. The initial temperature of the core is uniform and equals 10 K. Outside the cloud, the matter is also set to 10 K. The adiabatic index is set to $\gamma=7/5$.  We use the Rosseland $\kappa_\mathrm{R}$ and Planck $\kappa_\mathrm{P}$ mean opacities derived in \cite{Semenov_et_al_2003}. At temperature $>1000$ K, we impose $\kappa_\mathrm{P} =\kappa_\mathrm{R} = 0.01$ cm$^2$ g$^{-1}$, in order to limit the grain evaporation effect and to account for an inertia of the evaporation. We impose an initial subsonic  turbulent  velocity  dispersion which follows a Kolmogorov power spectrum $P(k)\propto k^{-5/3}$, where the phases are randomly sorted in the Fourier space.  Only one realization is explored in this study. The turbulence is not artificially sustained but, does not really decay as the gravitational time is typically shorter than the crossing time. The kinetic energy power spectrum peak roughly corresponds to the box size. 
The ratio of the turbulent to gravitational energies is given by $\alpha_\mathrm{turb}$.  
We do not include explicitly rotation but the turbulent field contains
angular momentum (both local and global). Note that our initial conditions exhibits a relatively flat density profile, which should favor fragmentation \citep{Girichidis_et_al_2011}. 

The magnetic intensity is set by the parameter $\mu = (M/\Phi)/(M/\Phi)_\mathrm{crit}$, which represents the value of mass-to-flux over critical mass-to-flux ratio \citep{ Mouschovias_Spitzer_1976,Hennebelle_et_al_2011}. In this Letter, we explore 3 magnetization degrees: $\mu=130$ corresponding to a  weak initial magnetic field, and $\mu=5$ and $\mu=2$ which are close to the observed values \citep[e.g.][]{Falgarone_et_al_2008}. We also investigate an almost spherical case with no magnetic field and no turbulence (model SPHYDRO) to serve as simple reference with respect to which the other simulations can be compared. All the simulations parameters are summarized in Table \ref{condinit}. Future work will imply further investigations on the effects of turbulence or rotation, but this goes beyond the scope of this Letter.

\subsection{Numerical method}

We use the adaptive mesh refinement code {\ttfamily{RAMSES }}\rm  \citep{Teyssier_2002}, which integrates the self-consistent equations of RMHD using ideal MHD  \citep{Fromang_2006,Teyssier_2006}, and flux limited diffusion for  RHD \citep{Commercon_2011a}. 

We impose a refinement criterion $N_\mathrm{J}=10$, which insures that the local Jeans length is resolved by at least 10 cells. The initial grid contains $256^3$ cells and we use 10 levels of refinement for an effective resolution of $262144^3$ and a minimum grid size of $\sim2.16$ AU. 
We apply periodic boundary conditions for hydrodynamics and gravity, and we impose a temperature of 10 K for the radiation at the edge of the box. 

\begin{table*}[t]
\begin{center}
\caption{Models parameters.\label{condinit}}
\begin{tabular}{cccccc}
\tableline\tableline
Model & $\mu$ & $\alpha_\mathrm{turb}$ & $\Delta$x$_\mathrm{min}$ (AU) & Coarse grid & $t_0$ (Myr) \\
\tableline
SPHYDRO & $\infty$          & $\sim 10^{-5}$& 2.16   & $128^3$ & 0.4786 \\
MU130  & $\sim136$  & $\sim 0.2$ & 2.16 & $256^3$ & 0.4935\\
MU5       & $\sim5.3$   & $\sim 0.2$ & 2.16 & $256^3$ & 0.5397 \\
MU2       & $\sim2.3$   & $\sim 0.2$ & 2.16 & $256^3$ & 0.5982\\
\tableline
\end{tabular}
\end{center}
\end{table*}

\begin{figure*}[t]
\includegraphics[scale=0.6]{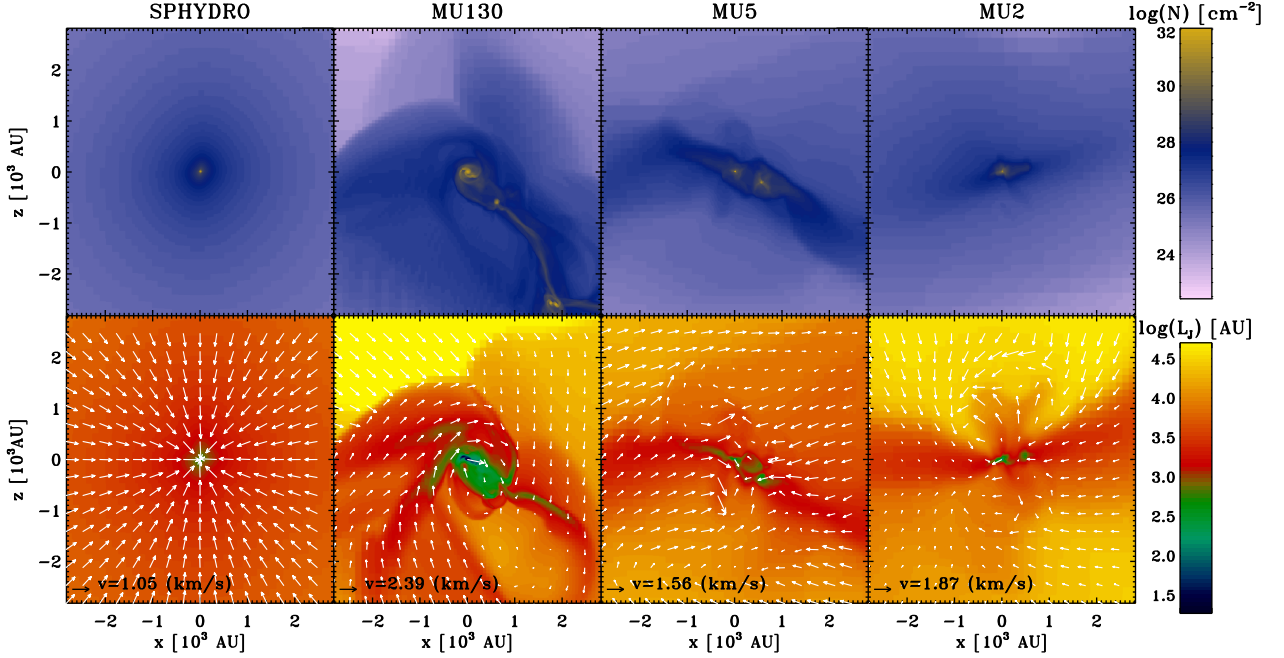}
\caption{{\it{Top:}} Column density maps integrated in the $y$-direction for the four models: SPHYDRO at time $\sim t_0 + 2.6$ kyr, MU130 at time $\sim t_0 + 25.4$ kyr,  MU5 at time $\sim t_0 + 7.5$ kyr,  and MU2 at time $\sim t_0 + 7.3$ kyr. {\it{Bottom:}} Local Jeans length and velocity field cut in the $xz$-plane for the same calculations and at the same time as in the upper row.}
\label{col_dens}
\end{figure*}

\section{Results}

\subsection{Qualitative description}

In this section, we qualitatively describe the results of the four calculations. Calculations are   synchronized at the time $t_0$ when the maximum level of refinement is reached (see table \ref{condinit}). The subsequent evolution after $t_0$ strongly depends on the physical conditions (rotation, outflows, etc...). At $t_0$, the first hydrostatic core \citep[FHSC, ][]{Larson_1969} is forming. As expected, we note that the stronger the magnetic field is, the later the collapse occurs because magnetic fields ``dilutes"  gravity. In Figure \ref{col_dens}, column density and local Jeans length maps of the four calculations are shown in the $xz$-plane. The boxes are centered at the maximum density of the total computational domain. 
In the SPHYDRO model, only one central fragment is formed and the collapse is nearly spherical. The mass of the fragment, i.e. where $\rho> 10^{11}$ cm$^{-3}$, at time $\sim t_0 + 2.6$ kyr is $\sim 0.2$ M$_\odot$. The integrated mass of the envelope from the fragment which is stable against fragmentation, i.e. $M_{int}/M_\mathrm{J} > 1$, is $\sim 30$ M$_\sun$ and the accretion rate  is  $\sim 10^{-4}$ M$_\odot$ yr$^{-1}$. 

In the MU130 model, the additional turbulent support clearly favors fragmentation (6 fragments are formed at that time) over a region of 2000-4000 AU. The mean separation between the fragments is about 1000 AU and they are distributed at this early time along  a filamentary structure.  This separation corresponds to the typical Jeans length associated with the region surrounding each fragment.  
 The overall accretion rate on the fragments is relatively low, $\sim 10^{-5}$ M$_\odot$ yr$^{-1}$, and consistent with the ones obtained in the low-mass star formation framework.  The  mean mass of the fragments is about $0.2$ M$_\odot$ ($\sim 1.2$ M$_\odot$ in total) and corresponds to the local Jeans mass.

In the MU5 model, the core has fragmented into two main objects, one being formed by the merger of secondary fragments. 
The fragmentation zone has the same filamentary morphology as in the MU130 model, but the extent is smaller. The angular momentum is more efficiently transported than in the MU130 model because of the stronger magnetic field \citep{Hennebelle_et_al_2011}. The region where rotation can support the collapsing cloud is thus smaller. The local Jeans length raises substantially indicating that the core is less prone to fragment. The accretion rate in this model is $\sim 10^{-4.2}$ M$_\odot$ yr$^{-1}$. The mass within each fragment is $\sim 0.25$ M$_\sun$.

\begin{figure}[t]
\includegraphics[scale=.55]{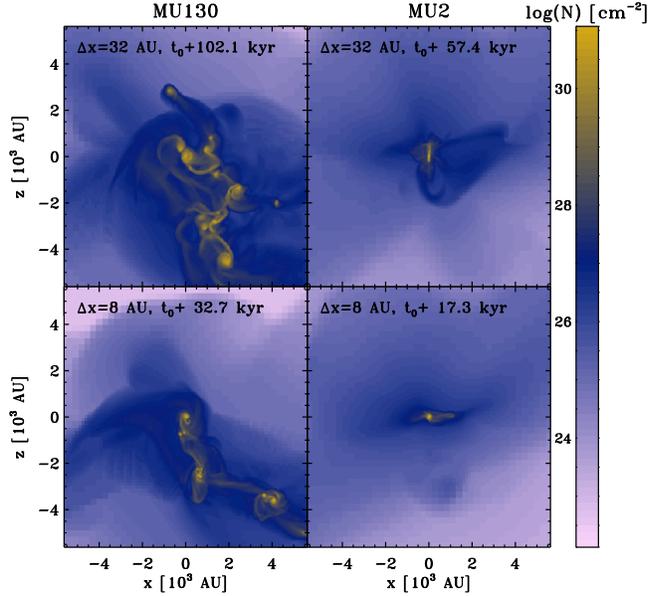}
\caption{Low resolution runs for the MU130 and MU2 models.}
\label{fig2}
\end{figure}

In the MU2 model, there is only one fragment which drives a low velocity outflow ($v_\mathrm{out}\sim 2$ km s$^{-1}$) of extent $\sim$ 1500 AU. 
 The accretion rate is $\sim 10^{-4.1}$ M$_\odot$ yr$^{-1}$. The local Jeans length is much larger than in the MU130 and MU5 models and shows some typical features of magnetized dense core collapse. The region of small Jeans length (L$_\mathrm{J} < 3000$ AU) corresponds to the pseudo-disk that has formed perpendicular to the magnetic field lines at larger scales. The feature that appears in the vertical direction  corresponds to the outflow that has been launched around the central fragment. As radiation preferentially escapes perpendicular to the pseudo-disk where the optical depth is smaller, the outflow region heats up and the local Jeans length raises. At time $\sim t_0 + 7.3$ kyr, the mass of the fragment is $\sim 0.25$ M$_\sun$ and the associated Jeans stable mass integrated through the envelope is $\sim 10$ M$_\sun$ and extends to a radius of $\sim 20 000$ AU. By contrast in the MU5 model, the stable Jeans mass associated to each fragment is $\sim 1.2$ M$_\sun$ and extends to a radius of $\sim 800$ AU. 
We note that we have performed lower resolution calculations ($\Delta x_\mathrm{min}\sim 32$ AU and $\Delta x_\mathrm{min}\sim 8$ AU) which do not show any fragmentation at latter times, while similar MU130 simulations show a lot of fragments (see fig. \ref{fig2}). At times given in fig. \ref{fig2}, the mass contained in the fragments  with $\Delta x_\mathrm{min}\sim 32$ AU  is $7.1\sim $ M$_\odot$ for the MU130 model and $\sim1.2$ M$_\odot$ for MU2 (respectively $\sim0.6$ M$_\odot$ and $\sim2.2$ M$_\odot$ with $\Delta x_\mathrm{min}\sim 8$ AU.)

\subsection{Quantitative analysis of the SPHYDRO model}

\begin{figure*}[t]
\includegraphics[scale=1]{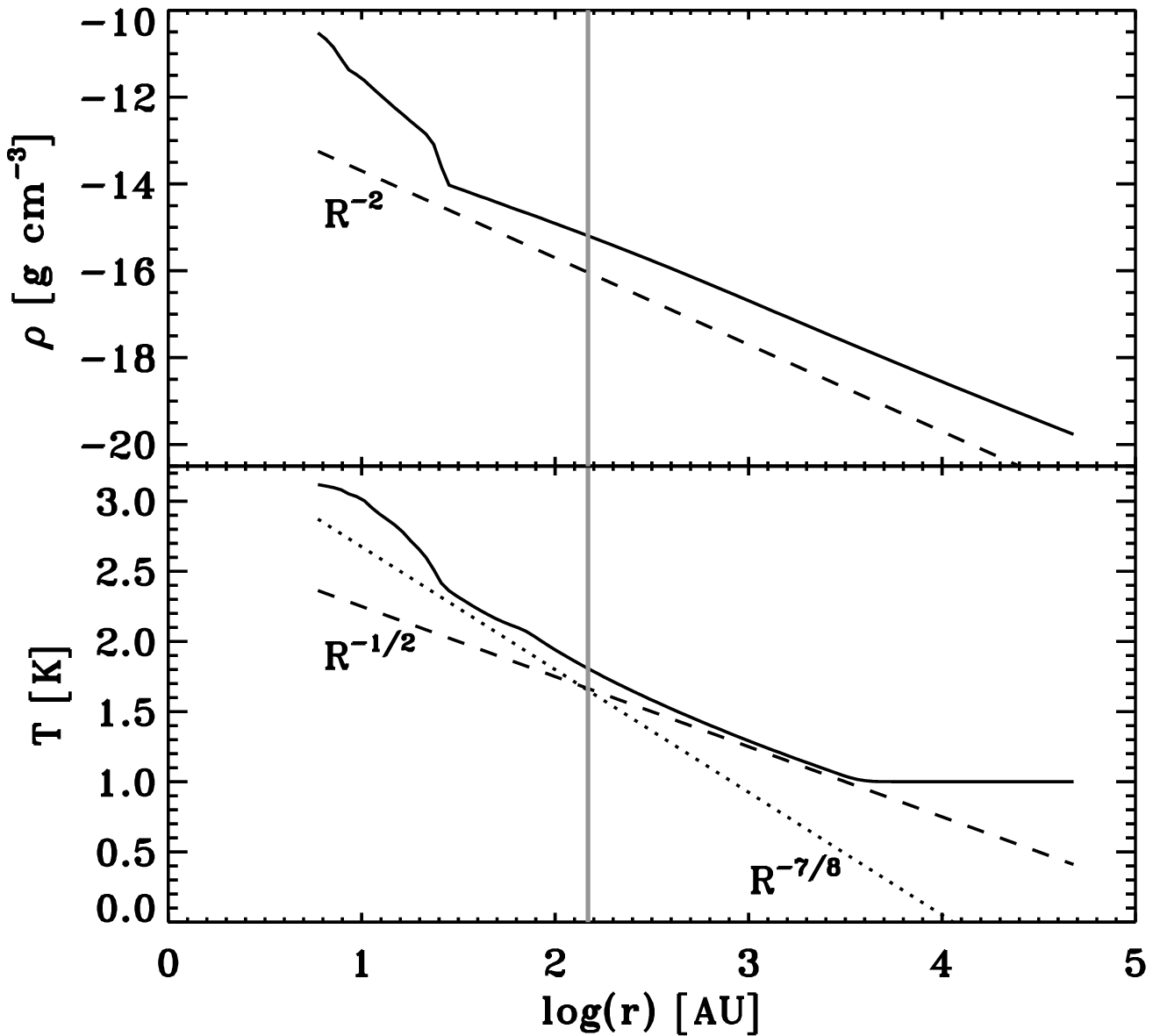}
\caption{Radial profiles of density and temperature for the SPHYDRO model at the same time as in figure \ref{col_dens}. The grey vertical lines indicates the radius at which $\tau=1$.}
\label{profiles}
\end{figure*}

In this section, we focus on the thermal behavior observed in the SPHYDRO model, i.e. without turbulence and magnetic field. Figure \ref{profiles} shows radial profiles of density and temperature at time $\sim t_0 + 2.6$ kyr . The density profile exhibits the classical $R^{-2}$ slope in the envelope \citep{Larson_1969,Penston_1969,Shu_1977}. Unlike the low mass star formation case, for which $T\propto R^{-0.5}$ in the optically thin envelope, we find that the temperature profile in the preshock region ahead of the FHSC is steeper. The optically thick region extends up to a radius of $\sim 130$ AU  (vertical grey line) much larger than the FHSC radius $R_\mathrm{fc} \sim 20$ AU. The radiation is thus trapped in an optically thick bubble where the infalling gas can be efficiently heated and the Jeans length raises efficiently. Compared to the low-mass star case, the accretion rate and thus the post-accretion shock temperature are  larger ($\sim$250 K compared to $\sim$70 K). Since  $\kappa_\mathrm{R}\propto T^2$ for temperature $<100$ K, this explains the larger optical depth found towards higher masses.
The temperature profile is well fitted in the optically thick region with $T\propto R^{-7/8}$ (dotted line). Such a profile is obtained assuming that the rate of change in  kinetic energy, $\sim G M\dot{M}/R$ equals the radiative flux, $\sim R^2 c/(3\kappa_\mathrm{R}\rho) a_\mathrm{R}T^4/R$ (with $\kappa_\mathrm{R}$ constant for temperature ranging from 100 K to 1000 K). At larger radii the optically thin behavior, $T\propto R^{-0.5}$ (dashed line), is recovered after a transition regime,  until the equilibrium temperature of 10 K is reached.

\subsection{Comparison with the other models}

\begin{figure*}[t]
\includegraphics[scale=.6]{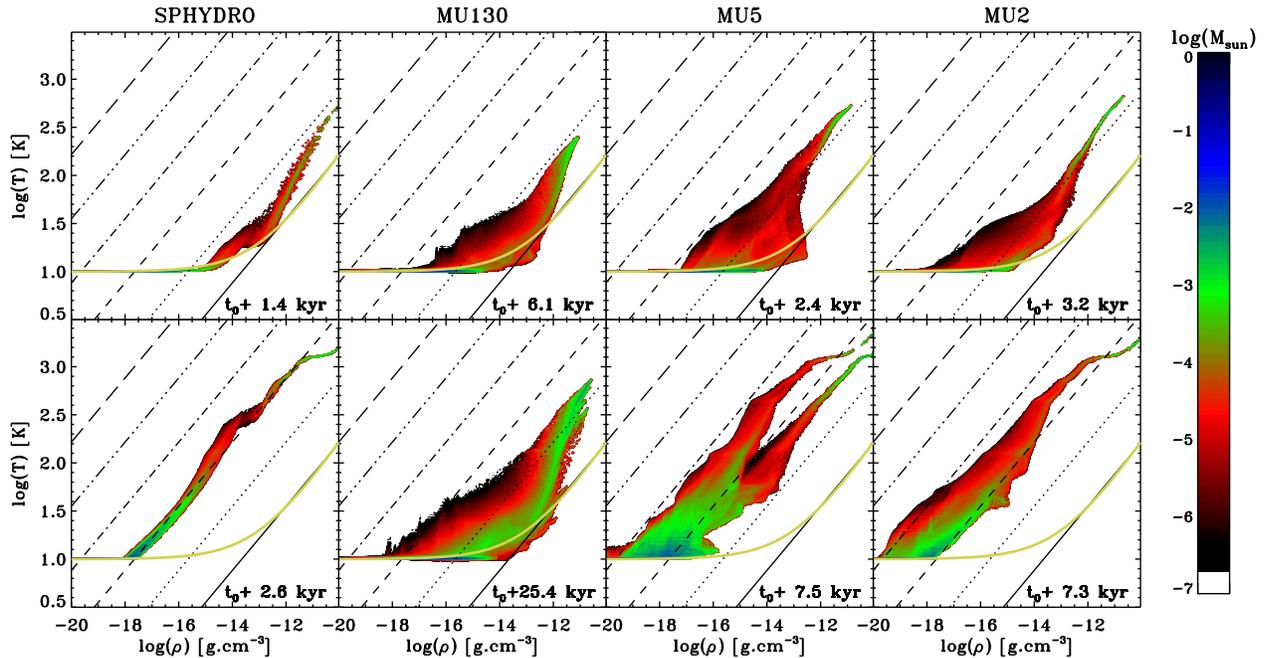}
\caption{Temperature-density distribution for the four models.  The top row corresponds to snapshots just before grain evaporation for the SPHYDRO, MU5, and MU2 models, and to time $t_0 + 6$ kyr for the MU130 model.  The bottom row corresponds to the same time as in figure \ref{col_dens}. The color coding indicates the mass in M$_\odot$ per bin of equal density and temperature. The black lines represent iso-Jeans mass curve, ranging from  $10^{-4}$ M$_\odot$ (bottom right line) to $10$ M$_\odot$ (top left dashed line). The yellow curve represents a classical barotropic EOS.}
\label{trho}
\end{figure*}

Figure \ref{trho} shows the temperature distribution  as a function of the density at two times for the four models.  
To visualize the difference between our work and other studies, we overplot the barotropic law that is used in \cite{Hennebelle_et_al_2011}.  With a full RMHD model, we obtain a spread in temperature and Jeans mass which depends on the magnetization of the core that a barotropic EOS cannot reproduce. The SPHYDRO and MU2  models are strikingly similar in the heating efficiency, even though the spread in temperature is larger in the MU2 model because of the initial turbulence. The rotation in not large enough to support the collapsing core at larger radii because of the magnetic braking, and matter is accreted onto a single fragment of similar size  and accretion rate as in the SPHYDRO model. The accretion luminosity on the FHSC, $L_\mathrm{acc}=GM_\mathrm{fc}\dot{M}/R_\mathrm{fc}$, 
is thus about the same. 
In the MU130 model, the accretion rate is much lower and the contraction of the FHSC slower because of lower temperature. As a consequence, the typical Jeans mass  is about 2 orders of magnitude lower than in the SPHYDRO and MU2 models. As in other models \citep{Krumholz_et_al_2009,Peters_et_al_2011, Hennebelle_et_al_2011}, the MU130 model shows  fragmentation in several objects even at time $t_0 + 6$ kyr, where three objects formed  with separation of 3000-5000 AU.  
As the magnetic braking is very small here, the radius
at which the collapse is stopped and where fragmentation is taking place,
is thus much larger than the FHSC radius, and the accretion luminosity
negligible.  Even with a fragment undergoing  second collapse, the radiative feedback would not be efficient at a distance of $d\sim3000$ AU to suppress fragmentation. For an optically thin envelope, the temperature indeed scales as $T\sim(L/(4\pi\sigma d^2))^{1/4}$, where $L$ is the accretion luminosity on the protostar, $L=GM_\star\dot{M}/R_\star$
 ($\sim13$K at 3000 AU with $\dot{M}=1\times10^{-5}$ M$_\odot/$yr, $M_\star=0.2$ M$_\odot$, and $R_\star\sim5$ R$_\odot$). Finally, the MU5 model shows an intermediate behavior. Although the heating owing to the accretion shock on the central fragment is already large, external regions have time to cool and to continue collapsing to form a secondary fragment.

\section{Towards massive star formation?}

\subsection{Discussion}

As first suggested by \cite{Commercon_2010} in the low mass star formation framework, we show that there is a strong {\it{interplay }}\rm between magnetic field and radiative transfer which may suppress initial fragmentation. Because of the rotation slowdown owing to the magnetic braking (most efficient at scales $< 200-300$ AU), the infall velocity on the fragment is then greater. For massive dense cores, the accretion is larger than for low mass ones and the amount of energy radiated away at the FHSC surface, the accretion luminosity, is thus much larger. 
Depending on the initial conditions, we find three totally different behaviors: 
\begin{itemize}
\item For a spherical massive dense core, without turbulence and magnetic field, we find that the core's envelope heats efficiently (SPHYDRO model), because the radius at which the accretion luminosity is released is small (FHSC radius). We also find that contrary to the low-mass regime, the optically thick regions is much more extended, up to $\sim 150$ AU ($\sim$10 AU for the low-mass case). Only one fragment is formed;
\item In the MU130 model, the magnetic braking is too low to transport outwards angular momentum produced by the initial turbulence. 7 fragments have formed by the end of the calculations. The fragmentation zone extends over a few thousands AU, at which accretion luminosity is first released through  a radiative shock. Disks with radius $\sim 100-200$ AU are also formed around the fragments, which give a second accretion shock at the disk edges. The infalling gas thus encounters several accretion shocks before being accreted onto several FHSCs, which lowers  the accretion luminosity. 
\item For highly magnetized and turbulent massive dense cores (MU2 model) we find very similar results to the SPHYDRO model, i.e. a single fragment, because of the magnetic braking that lowers the radius at which the accretion luminosity is released to the FHSC radius.
\item We find an intermediate behavior for lower magnetization degrees (MU5 model), where the core fragments into two objects of similar properties. The latter model may be the most realistic one, in accordance with observations (see hereafter).

\end{itemize}

\subsection{Astrophysical consequences}

Our results suggest that the {\it{combined }} effect of  magnetic fields and radiative transfer could control the early  fragmentation of the core and we speculate that this could lead 
either to the formation of isolated massive stars or  OB associations.  
The MU2 model gives raise to a single fragment with a large reservoir of mass $\sim 10$ M$_\odot$ stable against fragmentation. The strong magnetic field case is thus a preferred scenario for non-runaway massive stars (i.e. not ejected form a cluster) that are found in isolation. Our proposed scenario does not exclude  the formation of close massive binaries and OB associations. Fragmentation in massive binary system can possibly occur in more massive cores or in cores with different initial
conditions than ours (for example with stronger density fluctuations
initially). They could also form
during the second collapse phase as it has been shown in the low mass star formation framework \citep{Machida_et_al_2008}. OB associations can also form by global collapse of a giant molecular cloud containing several massive magnetized dense cores. 
In addition, the MU5 model produce two fragments, that are also associated with a relatively high Jeans mass reservoir. Contrary to previous lower resolution studies \citep{Krumholz_et_al_2007,Krumholz_et_al_2009,Peters_et_al_2011}, the secondary fragment is not produced by disc fragmentation, but rather by collapse along a filamentary structure. One can expect 
that this early fragmented system will also give rise to a close massive binary system following results of \cite{Bonnell_Bate_2005}.  

Recently  \cite{Girart_et_al_2009} report observations of a hot molecular core (HMC) in the massive star-forming region G3141 and conclude that the gravitational collapse of the HMC is controlled by magnetic field. They also observe a spin-down in the HMC which suggests that magnetic braking is acting and removing angular momentum. Last but not least, they infer a mass-to-flux over critical mass-to-flux ratio of 2.7 which corroborates our results. At later evolution stages,  \cite{Bestenlehner_et_al_2011} observe a very massive star of $\sim 150$ M$_\odot$ in apparent isolation from the massive cluster R136, which could be formed from a highly magnetized dense core rather than being ejected form the dense cluster. \cite{Bontemps_et_al_2010} report imaging of massive dense cores with high angular resolution. They find that fragmentation in massive core tends to lead to fewer high-mass fragments inconsistent with a pure gravoturbulent fragmentation (which would correspond to the MU130 model). Their results support our findings on the enhanced effect of the magnetic braking and radiative transfer when mass  increases.

\section{Conclusion and prospects}

In this letter, we propose a new mechanism to suppress initial fragmentation of highly magnetized massive dense cores and speculate that  it can lead to massive star formation. Our scenario differs form previous work, since it does not invoke stellar mergers as in the competitive accretion scenario \citep{Bonnell_et_al_2001}, nor low mass protostars' radiative feedback and high column density \citep{Krumholz_McKee_2008}, nor disk fragmentation \citep{Krumholz_et_al_2009,Peters_et_al_2011}. 
We investigate the early stages of the collapse and fragmentation of turbulent, massive, and magnetized dense cores with RMHD calculations. We show that the combined effect of magnetic braking and radiative transfer suppresses fragmentation in the case of a strong magnetic field as it has been shown in the low-mass star formation framework \citep{Commercon_2010,Tomida_et_al_2010}. Magnetic braking transports angular momentum outwards, and the accretion rate on the FHSC is thus larger.  As shown in \cite{Commercon_2011b}, all the infall kinetic energy is radiated away at the first core accretion shock and allows greater heating. The interplay between magnetic fields and radiative transfer is  independent from the initial conditions other than magnetic fields strength. In addition, this interplay can only be caught in self-consistent RMHD models with thermal (re-)emission and radiative shocks. 
This effect is expected to become stronger with appropriate physical models for the second collapse and second core formation and for protostellar evolution \citep[see e.g.][]{Krumholz_et_al_2007}. To reach the final stellar mass, the subsequent evolution of the forming protostars can then follow a disk accretion scenario \citep{Kuiper_et_al_2011,Seifried_et_al_2011}. Further work will imply  detailed fragmentation and resolution studies \citep[e.g.][]{Federrath_et_al_2011} and the  introduction of  sink particles to follow a longer dynamical range. We also neglect in this work  ionization and protostellar feedbacks which influence further time evolution \citep{Peters_et_al_2011, Cunningham_et_al_2011}.

As a conclusion, we speculate that highly magnetized dense cores are the seed of massive stars and good candidates for massive star forming regions.

\acknowledgments

Calculations have been performed on the THEO cluster at MPIA and on the JADE cluster at CINES. BC thanks Romain Teyssier and Henrik Beuther for useful discussions.  

%



\clearpage

\end{document}